\documentclass[aip,reprint]{revtex4-1}
\usepackage{graphicx}
\usepackage{dcolumn}
\usepackage{sidecap}
\usepackage[top=1.3cm, bottom=1.3cm, left=1.5cm, right=1.5cm]{geometry}
\usepackage[centerlast,font=small]{caption}
\DeclareCaptionType[name={Supplementary Figure}]{Sfigure}

\begin{document}

\title{Networks of silicon nanowires: a large-scale atomistic electronic
structure analysis}

\author{\"{U}mit Kele\c{s}}
\affiliation{Department of Physics, Bilkent University, Bilkent,
Ankara 06800, Turkey}
\author{Bartosz Liedke}
\author{Karl-Heinz Heinig}
\affiliation{Helmholtz-Zentrum Dresden~-~Rossendorf, Bautzner
Landstr. 400, 01328 Dresden, Germany}
\author{Ceyhun Bulutay}
\email{bulutay@fen.bilkent.edu.tr} \affiliation{Department of
Physics, Bilkent University, Bilkent, Ankara 06800, Turkey}

\date{\today}

\begin{abstract}
Networks of silicon nanowires possess intriguing electronic
properties surpassing the predictions based on quantum confinement
of individual nanowires. Employing large-scale atomistic
pseudopotential computations, as yet unexplored branched
nanostructures are investigated in the subsystem level, as well as
in full assembly. The end product is a simple but versatile
expression for the bandgap and band edge alignments of
multiply-crossing Si nanowires for various diameters, number of
crossings, and wire orientations. Further progress along this line
can potentially topple the bottom-up approach for Si nanowire
networks to a top-down design by starting with functionality and
leading to an enabling structure.
\end{abstract}


\maketitle

The sway of silicon technology on the industrial-scale fabrication
generally fosters developments within the same material paradigm
of silicon and its native oxide. In this respect, under the
pressing demands on functionality and reconfigurability,
silicon-based nano-networks (SiNets) with their added dimensional
and architectural degrees of freedom will undoubtedly be embraced
by the semiconductor community.\cite{lu2008}
The looming appearance of SiNets hinges upon the advancements made
on the synthesis of silicon nanowires (NWs) within the past
decade.\cite{morales1998,holmes2000,cui2001wire,ma2003,wu2004} En
route, branched nanocrystals,\cite{manna2003}
branched\cite{wang2004,jun2010,jiang2011} and
tree-like\cite{dick2004} NWs were realized, followed by the
connection of these individual branched nanostructures into
large-scale nanowire networks,\cite{zhou2005wo,dick2006} in some
cases using other semiconductors. In particular to SiNets, recently
several synthesis procedures have been accomplished which are
employed in the fabrication of thermoelectric
devices,\cite{totaro2012} biosensors,\cite{serre2013} and
photodetectors.\cite{mulazimoglu2013}

Notwithstanding, there appears to be a very limited understanding
of how to tune the electronic properties of NW networks, which is
further compounded by the electrical contact design requirements
for proper band alignments. As the encompassing gist of such
technical issues, essentially we need to know how will the
promising NW utilities be taken over to networks, and will
networks reveal even new features? These questions form the
aspiration of our work. In no doubt, only after having a solid
understanding of the underlying electronic properties, can one
suggest optimum SiNet morphologies tailored to specific
functionalities.

Experimental difficulties on determining the electronic properties
of nanostructures call for realistic computational tools with
predictive capabilities. For Si NWs, many electronic structure
calculations already exist in the literature.\cite{zhao2004,
bruno2007,yan2007,huang2008,rurali2007,ng2007,scheel2005,niquet2006,kim2011}
A comprehensive review on the theoretical investigations about Si
NWs is given by Rurali.\cite{ruralirev} In contrast to the single
Si NWs, hitherto, only a few theoretical studies have been carried
out for the electronic properties of branched Si NWs, and
virtually none on SiNets. Menon \textit{et al.}\cite{menon2007}
investigated branched pristine Si NWs, whereas actual grown NWs
have always passivated surfaces. Avramov \textit{et
al.}\cite{avramov2007nano} considered some very small size
flower-like Si nanocrystals rather than branched Si NWs. The lack
of more realistic theoretical attempts in such an experimentally
attractive area can be explained by the fact that even the
smallest branched systems, including surface passivation, contains
$\sim$10$^{3}$ to 10$^{4}$ atoms in the computational supercell.
The comprehensive study of such structures is one of the
prevailing challenges for the first-principles methods due to
drastically increased computational load which inevitably invites
more feasible semiempirical techniques.

The aim of this study is to lay the groundwork in the level of
single-particle semiempirical atomistic
pseudopotentials\cite{bester} for SiNets embedded in SiO$_{2}$
within a restricted energy range around the bandgap. For this
purpose, first, we consider Si NWs oriented in $\langle 100\rangle
$, $\langle 110\rangle $, $\langle 111\rangle$, and $\langle
112\rangle $ crystalline directions. We calculate their energy
gaps as well as valence and conduction band edge alignments with
respect to bulk Si as a function of wire diameter. This is
followed by a detailed analysis of two- and three-dimensional
SiNets to establish an understanding of their electronic
properties. Comprehensive results are consolidated into a general
expression which provides a simple way to estimate the electronic
properties of SiNets.

In our computational framework, the semiempirical
pseudopotential-based atomistic Hamiltonian is solved using an
expansion basis formed by the linear combination of bulk bands
(LCBB) of the constituents of the nanostructure, i.e., Si and
SiO$_2$.\cite{lcbb1997,lcbb1999} While non-self-consistent in
nature, semiempirical techniques have the benefit that the
calculated bandgaps of nanostructures inherently agree with the
experimental values.\cite{bester,kim2011} The surface passivation
is provided by embedding the NW structures into an
\emph{artificial} wide bandgap host matrix which is meant to
represent silica.\cite{bulutay2007} In particular, the embedding
matrix has the same band edge line up and dielectric constant as
silica, but it is lattice-matched with the diamond structure of
Si. (For details, see the supplementary
material.\cite{suppl_matl}) Though missing surface relaxation and
strain effects, the competence of our method has been validated,
in the context of embedded Si and Ge nanocrystals, confronting
with experimental data for the linear\cite{bulutay2007} and
third-order nonlinear optical
properties,\cite{yildirim2008,imakita2009} and the
quantum-confined Stark effect.\cite{bulutay2010}

\begin{SCfigure*} [10][ht]
\caption{ \small Bandgap energies as a function of diameter for $[100]$, $[112]$, $%
[110]$, and $[111]$-oriented Si NWs. Our results for
oxide-passivated NWs (solid lines with filled circles) are
compared with data of experimental\cite{ma2003} and theoretical
results of H-passivated NWs including empirical pseudopotential
method,\cite{kim2011} semiempirical tight
binding,\cite{niquet2006} and density functional theory
calculations
correcting bandgaps with GW approximation,\cite%
{zhao2004,yan2007,bruno2007,huang2008} hybrid functionals\cite%
{rurali2007,ng2007} or scissors operation.\cite{scheel2005}
\newline \newline \newline \newline \newline \newline \newline } \label{lit}
\includegraphics[width=0.65\textwidth]{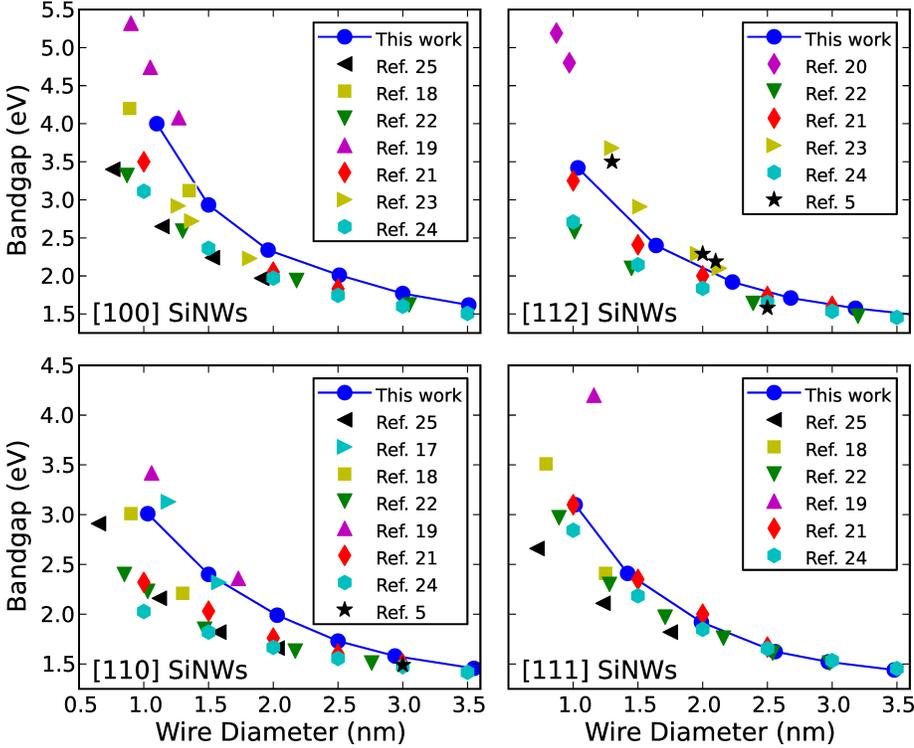}
\end{SCfigure*}

To set the stage for SiNets, first we establish the single NW
case. For oxide-passivated Si NWs aligned along the $\langle
100\rangle $, $\langle 110\rangle $, $\langle 111\rangle $, and
$\langle 112\rangle $ crystalline orientations, we report the
effective electronic bandgaps, regardless of whether it is direct
or indirect. In Fig.~\ref{lit}, our NW results are compared with a
compilation of some representative experimental and theoretical
data for wire diameters in the range of 0.5~nm to 3.5~nm. Note
that a strict comparison will not be meaningful as the Si NWs
considered here are embedded in SiO$_{2}$, whereas the literature
values are for H-passivated Si NWs. In general terms, oxide
passivation is observed to show similar trends like that of
H-passivation.

\begin{figure}[b]
\includegraphics[width=0.45\textwidth]{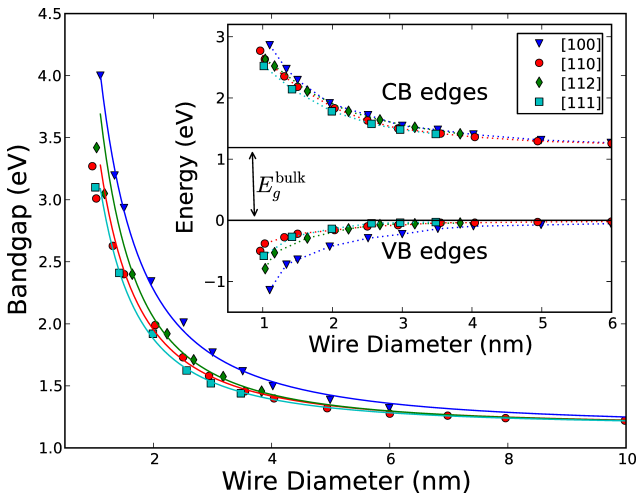}
\caption{ \small Bandgap energy of oxide-passivated Si NWs as a
function of diameter. The values are fitted with a $C
d^{-\protect\alpha}$ form. Inset shows the variation of valence
and conduction band edges. The dashed lines of the inset are just
guides to the eyes.} \label{wire}
\end{figure}

In Fig.~\ref{wire}, we display our bandgap results for different
NW orientations. For all directions, the gaps decrease
asymptotically towards the bulk Si value with increasing wire
diameter, reflecting the reduction of the quantum confinement
effect. In this figure, dependence of the bandgap $E_{g}$ on the
wire diameter is described by
\begin{equation} \label{fitwire}
E_{g}=E_{g}^{\mathrm{bulk}}+Cd^{-\alpha }\,,
\end{equation}
as proposed according to the effective mass
approximation.\cite{delerue1993} In this expression, $d$ is the
diameter of the wire, $C$ and $\alpha $ are fitting parameters,
and $E_{g}^{\mathrm{bulk}}$=1.17~eV is the experimental bulk Si
bandgap value. The fitted \{$C$, $\alpha $\} parameters are listed
in Table \ref{tbl:cpar}. In the fitting procedure, we include data
points for diameters above $\sim$1.3~nm. This is based on our
observation that in the excluded strong confinement regime a
different physical mechanism sets in. Namely, the wave function
penetrates into the oxide matrix, thereby experiencing a larger
effective diameter.\cite{scheel2005} For this reason, in
Fig.~\ref{wire} the data points of smaller diameters somewhat
deviate from the fitting curves.

We can note that the hallmark of the quantum confinement within
the effective mass approximation is the $1/d^2$ scaling of the
single-particle state energies.\cite{brus83} The same behavior
prevails even when valence band coupling and conduction band
off-$\Gamma$ minima are taken to account in the underlying band
structure.\cite{takagahara92} On the other hand, in our analysis
the exponent, $\alpha$ significantly deviates from 2 to values in
the range 1.57 to 1.76 depending on the direction (Table
\ref{tbl:cpar}). Discrepancy stems from the lack of atomistic
potentials in the former that relies solely on the effective mass
and the kinetic energy of the carriers.\cite{brus83} Our assertion
is that an atomistic treatment becomes crucial even close to the
band edge energies.

In the inset of Fig.~\ref{wire}, we plot the variation of valence
band (VB) and conduction band (CB) edges as a function of wire
diameter for different wire orientations. Indicating the bulk Si
band edges with horizontal solid lines, this figure also
illustrates the alignments of band edges of Si NWs with respect to
bulk Si for increasing diameters. We find out that the band edge
energies have the same functional dependence with respect to NW
diameter just like the bandgaps, as in Eq.~(\ref{fitwire}). Hence,
setting the bulk VB maximum of Si to zero, VB and CB edge energies
can also be described by
$E_{\mathrm{VBE}}=C_{\mathrm{VBE}}d^{-\alpha _{%
\mathrm{VBE}}}$ and $E_{\mathrm{CBE}}=E_g^{\mathrm{bulk}}+C_{%
\mathrm{CBE}}d^{-\alpha _{\mathrm{CBE}}}$, where \{$C_{%
\mathrm{VBE}},\alpha _{\mathrm{VBE}}$\} and \{$C_{\mathrm{CBE}},\alpha _{%
\mathrm{CBE}}$\} are fitted for the data points given in the inset
of Fig.~\ref{wire}, and listed in Table \ref{tbl:cpar}. Notably,
$\alpha_{\mathrm{CBE}}$ behave similar to the bandgap exponents
$\alpha$, with either one being relatively less sensitive to wire
directions. In striking contrast, $\alpha_{\mathrm{VBE}}$ displays
a curious dual character: $\langle111\rangle$ and
$\langle112\rangle$ wires have values around 2, whereas
$\langle100\rangle$ and $\langle110\rangle$ substantially deviate
from the quadratic behavior.\cite{edgenote} Having device
applications in mind, in the supplementary material, we also
provide the band offsets of Si NWs with respect to bulk Si and
SiO$_2$.\cite{suppl_matl}

\begin{table}[t]
\caption{ \small Wire orientation-dependent fitting parameters
associated with Eq.~(\ref{fitwire}) (also used for
Eq.~(\ref{fitgeneral})) for the main gap energy ($C$,~$\alpha$) as
well as for VB ($C_{\mathrm{VBE}}$,~$\alpha_{\mathrm{VBE}}$) and
CB ($C_{\mathrm{CBE}}$,~$\alpha_{\mathrm{CBE}}$) edge energies.
When diameters $d$ in Eqs.~(\ref{fitwire}) and (\ref{fitgeneral})
are in nm units, the energies come out in units of eV.}
\label{tbl:cpar}
\begin{tabular*}{\linewidth}{@{\extracolsep{\fill}}ccccc}
\hline \hline
& $\langle100\rangle$ & $\langle110\rangle$ & $\langle111\rangle$ & $%
\langle112\rangle$ \\ \hline
$C$ & 3.31 & 2.47 & 2.25 & 2.98 \\
$\alpha$ & 1.57 & 1.66 & 1.67 & 1.76 \\
$C_{\mathrm{VBE}}$ & -1.22 & -0.45 & -0.49 & -0.69 \\
$\alpha_{\mathrm{VBE}}$ & 1.64 & 1.66 & 1.94 & 1.98 \\
$C_{\mathrm{CBE}}$ & 2.11 & 2.07 & 1.74 & 2.25 \\
$\alpha_{\mathrm{CBE}}$ & 1.63 & 1.75 & 1.66 & 1.74 \\
\hline
\hline
\end{tabular*}
\end{table}

Regarding the H-passivated Si NWs, it is known that at a given
wire diameter up to around 3~nm, the bandgap follows the ordering
$E_{g}^{\langle 100\rangle }>E_{g}^{\langle 111\rangle }\sim
E_{g}^{\langle 112\rangle }>E_{g}^{\langle 110\rangle }$ with NW
orientation.\cite{ng2007,ruralirev} In our oxide-passivation case,
although $\langle 100\rangle $ Si NWs have the largest bandgap
energy as before, the ordering changes to $E_{g}^{\langle
100\rangle }>E_{g}^{\langle 112\rangle }>E_{g}^{\langle 110\rangle
}\sim E_{g}^{\langle 111\rangle }$ (see Fig.~\ref{wire}). This
ordering is robust under several different pseudopotential
parameterizations that we tried for the oxide matrix. Moreover,
this observation is in accord with the results of
Ref.~\citenum{bondi2011}, where the electronic structure of
oxide-sheathed Si NWs is compared with reference H-passivated Si
NWs and it is reported that the magnitude of $E_{g}$ shrinks for
$\langle 100\rangle$ and $\langle 111\rangle$-oriented SiNWs,
while it increases for both $\langle 110\rangle$ and $\langle
112\rangle$-oriented SiNWs. Given the fact that only one
calculation was provided in that work for each orientation, more
extensive first-principles calculations are required to extract
general trends on the ordering of bandgaps of oxidized Si NWs with
wire alignment.

Yan \textit{et al.}\cite{yan2007} and Niquet \textit{et
al.}\cite{niquet2006} give the VB and CB edges for H-passivated Si
NWs which agree with the general trends of our VB edge energies
(see the inset of Fig.~\ref{wire}). However, for the CB edge
energies of $\langle 112\rangle$ and $\langle 110\rangle$ Si NWs,
they report a smaller variation with diameter (even less than
0.5~meV for $\langle 110\rangle$ Si NWs), while in our case, those
values change as much as the CB edges of $\langle 100\rangle$ and
$\langle 111\rangle$ Si NWs. This behavior of $\langle 112\rangle$
and $\langle 110\rangle$ Si NWs gives rise to the distinct bandgap
anisotropy of our oxide-passivated Si NWs.

\begin{figure}[t]
\includegraphics[width=0.45\textwidth]{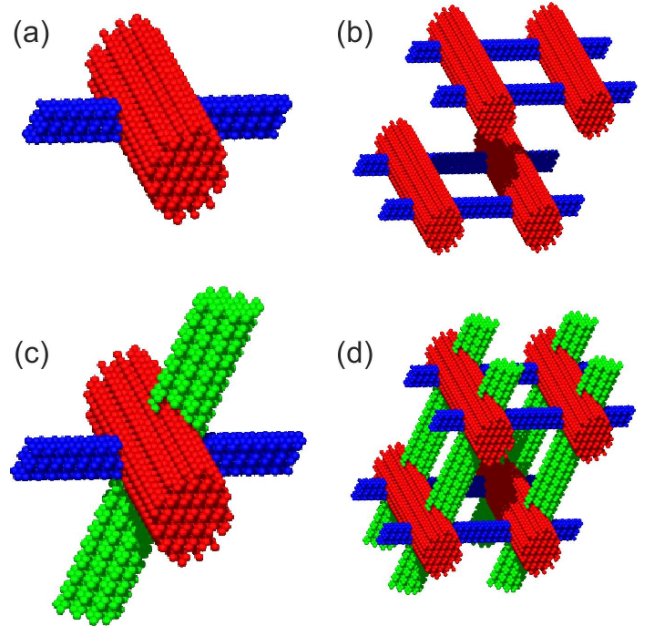}
\caption{ \small The computational supercells contain (a) two- or
(c) three-wire crossings to build up (b) two- or (d)
three-dimensional continuous SiNets, respectively. The wire
orientations are in the family of $\langle 110\rangle $
directions. For clarity, Si atoms are shown in different colors
for the crossing wires, and the matrix atoms are not
shown.}\label{nnet}
\end{figure}

Next, we consider the crossings of Si NWs as building blocks of
SiNets (see Fig.~\ref{nnet}). The branched Si NWs synthesized so
far have the tendency to grow in the $\langle $111$\rangle$
crystal directions with larger diameters
($>$~20~nm).\cite{wang2004,dick2004,jun2010} Whereas, at sub-10~nm
diameters the occurrence of $\langle $110$\rangle$ NW alignment
trumps over other directions.\cite{cui2001wire,wu2004,schmidt2005}
Henceforth, our results, referring to sub-10~nm diameter crossings
of Si NWs, are mainly quoted for the crossings of $\langle
110\rangle $-oriented NWs. Nevertheless, we have performed
calculations for crossings of $\langle 100\rangle $, $\langle
110\rangle $, $\langle 111\rangle $, and $\langle 112\rangle $
aligned NWs as well. Thus, our general conclusions are valid for
all directions unless stated otherwise.

As Figs.~\ref{nnet}(a) and (c) show, our computational supercells
contain crossings of two or three NWs. Taking into account the
periodic boundary conditions, the calculations are performed for
regular arrays of crossings (Figs.~\ref{nnet}(b) and (d)). These
structures are again embedded into the oxide matrix. We note that
in comparison to single NW calculations, network supercells
require much more atoms. For instance, in the supercell of a
crossing of three 3~nm-thick NWs, the supercell consists of 3694
Si and 7970 matrix atoms, respectively. We consider unrelaxed
crossings of Si NWs, i.e. no changes of crossing morphologies by
interface energy minimization are taken into account. Although,
the realistic crossings have some reconstructions, the electron
microscopy images show that the intercrossing regions are still
very close to the ideal unrelaxed
case.\cite{wang2004,jun2010,jiang2011}

\begin{table}[tbp]
\caption{ \small Bandgap energies for Si NWs (No.~1, 4, 7),
two-wire (No.~2, 5, 8, 9) and three-wire crossings (No.~3, 6,
10-14). Non-crossing bandgap value for each case is quoted in
brackets. }
\label{tbl:juncdata}
\begin{tabular*}{\linewidth}{@{\extracolsep{\fill}}cccccc}
\hline \hline & \multicolumn{3}{c}{Wire diameters}  \\
\cline{2-4}  No. & $d_1$ (nm) & $d_2$ (nm) & $d_3$ (nm) &
\multicolumn{2}{c}{$E_g$ (eV)}
\\ \cline{1-6}
1 & 1.50 & - & - & 2.40 & ~- \\
2 & 1.50 & 1.50 & - & 2.18 & ~[2.40] \\
3 & 1.50 & 1.50 & 1.50 & 2.11 & ~[2.40] \\
4 & 2.03 & - & - & 1.99 & ~- \\
5 & 2.03 & 2.03 & - & 1.84 & ~[1.99] \\
6 & 2.03 & 2.03 & 2.03 & 1.73 & ~[1.99]\\
7 & 2.94 & - & - & 1.57 & ~- \\
8 & 1.50 & 2.03 & - & 1.96 & ~[1.99]\\
9 & 1.50 & 2.94 & - & 1.57 & ~[1.57]\\
10 & 1.50 & 1.50 & 2.03 & 1.85 & ~[1.99] \\
11 & 1.50 & 2.03 & 2.03 & 1.81 & ~[1.99]\\
12 & 1.50 & 1.50 & 2.94 & 1.57 & ~[1.57]\\
13 & 1.50 & 2.03 & 2.94 & 1.56 & ~[1.57]\\
14 & 2.03 & 2.03 & 2.94 & 1.55 & ~[1.57]\\
\hline \hline
\end{tabular*}
\end{table}

The results for various combinations are summarized in
Table~\ref{tbl:juncdata}. Initially, to unambiguously address the
effect of crossing, we check the case when the participating NWs
do not cross each other (data shown in brackets in
Table~\ref{tbl:juncdata}), and are separated by at least 1~nm to
suppress their interactions. We observe that (i) the number of NWs
in the supercell does not alter the calculated bandgap when the
diameters are equal (No.~1-3, and 4-6), (ii) the bandgap is that
of the thickest NW when the diameters are distinct. On the other
hand, in the case of crossing NWs, the bandgap values reduce with
the increasing number of crossing wires (No.~1-3, and 4-6). This
is caused by the reduction in confinement due to the wave function
extensions into the branches. Here, in going from single NW to
two-wire crossing, a significant reduction in the bandgap occurs
that becomes not as pronounced when an additional third crossing
is introduced (No.~1-3). Other aspects of crossing can be
discussed referring No.~8 and 9: while the bandgap is determined
by the thicker NW in the latter, the thinner NW still influences a
marginal reduction in the former. Similar behavior is valid for
the crossings of three NWs that is, the bandgap is dominated by
the thickest NW while other NWs exert a reduction to the extent of
their diameters.

\begin{figure}[t]
\includegraphics[width=0.45\textwidth]{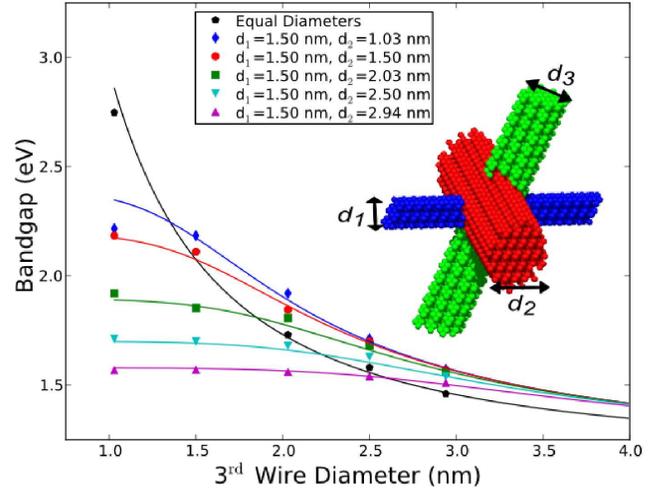}
\caption{Bandgap values for three-wire crossings. The calculated
data points are shown with the markers, the lines are obtained via
Eq.~(\ref{fitgeneral}) ($\beta$=5.5 is used,
see also Table \ref{tbl:cpar}). The crossing wire alignments are along $%
\langle110\rangle$ directions.} \label{junc}
\end{figure}

To estimate the bandgap values of SiNets, our observations on NW
crossings can be consolidated into a generalized form of
Eq.~(\ref{fitwire}). As our main result, we propose for the
bandgap $E_{g}$ of $N$ crossing wires, the expression
\begin{equation}\label{fitgeneral}
E_{g}=E_{g}^{\mathrm{bulk}}+C\left( \sum_{i=1}^{N}d_{i}^{\beta
}\right) ^{-\alpha /\beta }\,,
\end{equation}
where $E_{g}^{\mathrm{bulk}}$ is the bandgap of bulk Si and
the $d_{i}$ is the diameter of the NW indexed by $i$. Here $C$ and $%
\alpha $ are fitting parameters inherited from
Eq.~(\ref{fitwire}). Within the notion of generalized
mean,\cite{suppl_matl} the exponent $\beta$ governs the
contributions of each NW, namely the larger the parameter $\beta$,
the higher the contribution of the thickest NW to the bandgap.
Ultimately the specific value of $\beta$ is an outcome of the
material-dependent atomic potentials and
hence the quantum size effect. For a single NW ($N$=1), Eq.~(\ref%
{fitgeneral}) reduces to Eq.~(\ref{fitwire}) which suggest that
\{$C,\alpha$\} of single NWs as tabulated in Table~\ref{tbl:cpar}
can also be used for Eq.~(\ref{fitgeneral}). Regarding the
sensitivity to $\beta$, the estimated bandgap varies only by $\pm$
0.1~eV when $\beta$ changes in the range 4 to 7. Based on our
directional analysis, we suggest to use $\beta$=5.5 for $\langle
110\rangle $, $\langle 111\rangle $, and $\langle112\rangle$
crossings, whereas for $\langle 100\rangle$ crossings $\beta$=4
yields a better estimation. Figure~\ref{junc} shows the calculated
data points and corresponding plots of Eq.~(\ref{fitgeneral}) for
three-dimensional networks ($N$=3), for $\beta$=5.5. The overall
performance of Eq.~(\ref{fitgeneral}) is highly satisfactory with
the anticipated deviations for the very small diameters as we
discussed in the single NW case.

In order to estimate VB and CB offsets of networks with respect to
bulk Si, the form of Eq.~(\ref{fitgeneral}) can again be invoked
as done for the single NW case. Corresponding plots are given in
the the supplementary material~\cite{suppl_matl} employing
\{$C_{\mathrm{VBE}},\alpha _{\mathrm{VBE}}$\} and
\{$C_{\mathrm{CBE}},\alpha _{\mathrm{CBE}}$\} parameters of single
Si NWs (given in Table~\ref{tbl:cpar}) together with the same
$\beta$ values of bandgap estimation.


In conclusion, we computed the electronic bandgap energies of Si
NW structures embedded into silica using an atomistic
pseudopotential approach. First, we investigated the variation of
bandgap and band edge alignments as a function of wire diameter
for various orientations of Si NWs. Our results indicate a bandgap
anisotropy that differs from the H-passivated case. After
establishing the single-wire case, we extended our consideration
to the main subject of this paper, the two- and three-dimensional
SiNets. Based on a comprehensive analysis, we proposed an
expression to estimate the bandgap values of networks as a
function of crossing wire diameters. The form of the expression
should, in principle hold for other materials as well, to assist
bandgap engineering of NW networks. This expression can also be
used to calculate the valence and conduction band edge alignments
with respect to bulk Si. The semiempirical atomistic calculations
given in this work are for relatively large diameters. A
complimentary follow-up could be a first-principles investigation
for small-diameter networks to shed light especially on surface
chemistry and strain effects.
\\
\\

We would like to thank O\u{g}uz G\"ulseren for valuable
discussions. This work has been supported by The Scientific and
Technological Research Council of Turkey (T\"{U}B\.{I}TAK) with
Project No. 109R037 and German Federal Ministry of Education and
Research (BMBF) with Project No. TUR09240. \"{U}K acknowledges
Helmholtz-Zentrum Dresden~-~Rossendorf for supporting his visits in
Dresden.

%

\onecolumngrid

\newpage
\noindent\textbf{\Large Supplementary material for Networks of
silicon nanowires: a large-scale atomistic electronic structure
analysis}

\section{Some Technical Details on the Implementation}

In our implementation, the supercell size is set large enough to
prevent the interactions between the structures and their images
in neighboring cells (2~nm distance at least from surface to
surface for single NWs, and inter-crossing separation for SiNets).
Since the bandgap of Si NWs is rather insensitive to the
cross-section shape as long as the mean wire diameter is the
same,\cite{ng2007s} we carve out of the bulk Si an essentially
cylindrical shape NWs, and embed them into an oxide host matrix.
The resulting NWs are one-dimensional periodic structures. For
different NW directions, wire unit cell length $a_{\mathrm{wire}}$
has different values such as $a_{\langle100\rangle}=a$,
$a_{\langle110\rangle}= a/\sqrt{2}$, $a_{\langle111\rangle}=
a\sqrt{3}$, and $a_{\langle112\rangle}= a\sqrt{6}/2$, where
$a=$5.43~\AA~ is the Si lattice constant. We define the mean wire
diameter as the diameter of the cylinder of length
$a_{\mathrm{wire}}$ whose volume is the same as the average volume
$\Omega=N_{\mathrm{Si}} a^3/8$ occupied by the $N_{\mathrm{Si}}$
number of Si atoms in the wire unit cell.\cite{niquet2006s}  We
tested and observed that the effect of spin-orbit interaction
causes shifts of the band edge states by typically of at most
10~meV, and therefore we excluded this term from the Hamiltonian
in the interest of computational budget.

For the local empirical pseudopotentials of Si and SiO$_2$, we
employ the analytical expression given by Freidel $et$
$al.$\cite{friedel1989s} They use the following functional form
for the pseudopotential form factor at a given general wave number
$q$
\begin{equation}\label{pp}
V_{\mathrm{PP}}(q)=\frac{1}{2} \left[
 \tanh\left( \frac{a_5-q^2}{a_6} \right) + 1
\right] \frac{a_1(q^2-a_2)}{e^{a_3(q^2-a_4)}+1} .
\end{equation}
We employ their parameters for Si.\cite{friedel1989s} On the other
hand, SiO$_2$ is more subtle not only due to the presence of
oxygen but also due to its many polymorphs locally incompatible
with the diamond structure of Si. Therefore, we introduce an
artificial monoatomic wide bandgap material that reproduces the
refractive index and band alignments of SiO$_2$ with Si, however
with a diamond structure lattice-matched to Si to circumvent
further complications associated with strain.\cite{bulutay2007s}
As an improvement over our prior work,\cite{bulutay2007s} we
generate new pseudopotential parameters to represent the
artificial SiO$_2$ which satisfies the experimental band offsets
of bulk Si/SiO$_2$ interference having 4.4 and 3.4~eV for
valence~\cite{keister1999s} and conduction~\cite{himpsel1988s}
band edges, respectively. The resulted parameters in
Eq.~(\ref{pp}) are $a_1=69.625$, $a_2=2.614$, $a_3=0.0786$,
$a_4=-19.1433$, $a_5=5.99$, $a_6=0.335$. Here, the units are such
that the pseudopotential form factors come out in Rydberg energy
unit, and the wave number $q$ in the above expression should be in
atomic units (1/Bohr radius). In the LCBB basis
set\cite{lcbb1997s,lcbb1999s} we employ a large number of both
core and matrix Bloch states to attain high accuracy. This
produces not only core-derived states that we are interested in
this work, but also various matrix-derived interface states. We
filtered out the latter.

\section{Generalized Mean}
The estimation function of nanowire networks is constructed on the
notion of the generalized mean,\cite{abramowitz1965s} also known
as the power mean, to express our conclusion that thickest wire
dominates the bandgap while the narrower wires still contributing
marginally. The generalized $\beta$-mean of a set of values
($x_1,x_2,...,x_N$) is defined by the expression
\begin{equation}
M_{\beta} (x_1,x_2,...,x_N) = \left[  \frac{1}{N} \sum_{i=1}^{N}
x_i^{\beta} \right]^{1/\beta} .
\end{equation}
The special cases of generalized mean includes: minimum ($\beta
\rightarrow -\infty$), harmonic mean ($\beta=-1$), geometric mean
($\beta \rightarrow 0$), arithmetic mean ($\beta=1$), quadratic
mean ($\beta=2$), and maximum ($\beta \rightarrow +\infty$).

\section{Estimation expression for valence and conduction band offsets}

\subsection{For Si Nanowires}
We can give the band offsets of Si NWs with respect to bulk Si as
$\Delta_{\mathrm{VBO}}^{\mathrm{Si}}=
C_{\mathrm{VBE}}d^{-\alpha_{\mathrm{VBE}}}$ and
$\Delta_{\mathrm{CBO}}^{\mathrm{Si}}=
C_{\mathrm{CBE}}d^{-\alpha_{\mathrm{CBE}}}$.
Moreover, since we set the bulk Si/SiO$_{2}$ VB (CB) offset as $E_{\mathrm{%
VBO}}$=4.4~eV ($E_{\mathrm{CBO}}$=3.4~eV), we can calculate the
band offsets
of Si NWs with respect to silica by simply writing $\Delta _{\mathrm{VBO}%
}^{\mathrm{SiO}_{2}}=-E_{\mathrm{VBO}}-C_{\mathrm{VBE}}d^{-\alpha
_{\mathrm{VBE}}}$
 $\left(\Delta _{\mathrm{CBO}}^{\mathrm{SiO}_{2}}=E_{\mathrm{CBO}}-C_{\mathrm{CBE}%
}d^{-\alpha _{\mathrm{CBE}}}\right)$.

\subsection{For Si Nanowire Networks}

In Supplementary Fig.~\ref{vcbo}, we use
$\Delta^{\mathrm{Si}}_{\mathrm{VBO}}=C_{\mathrm{VBE}}\left(\sum_{i=1}^N
d_i^\beta\right)^{-\alpha_{\mathrm{VBE}} /\beta}$ and
$\Delta^{\mathrm{Si}}_{\mathrm{CBO}}=C_{\mathrm{CBE}}\left(\sum_{i=1}^N
d_i^\beta\right)^{-\alpha_{\mathrm{CBE}} /\beta}$ with single Si
NW band edge fitting parameters
\{$C_{\mathrm{VBE}}$,~$\alpha_{\mathrm{VBE}}$\} and
\{$C_{\mathrm{CBE}}$,~$\alpha_{\mathrm{CBE}}$\} from Table~I of
the main paper to estimate the band offsets of Si NW networks with
respect to bulk Si band edges.

\begin{Sfigure}[h]
\includegraphics[width=15.0cm]{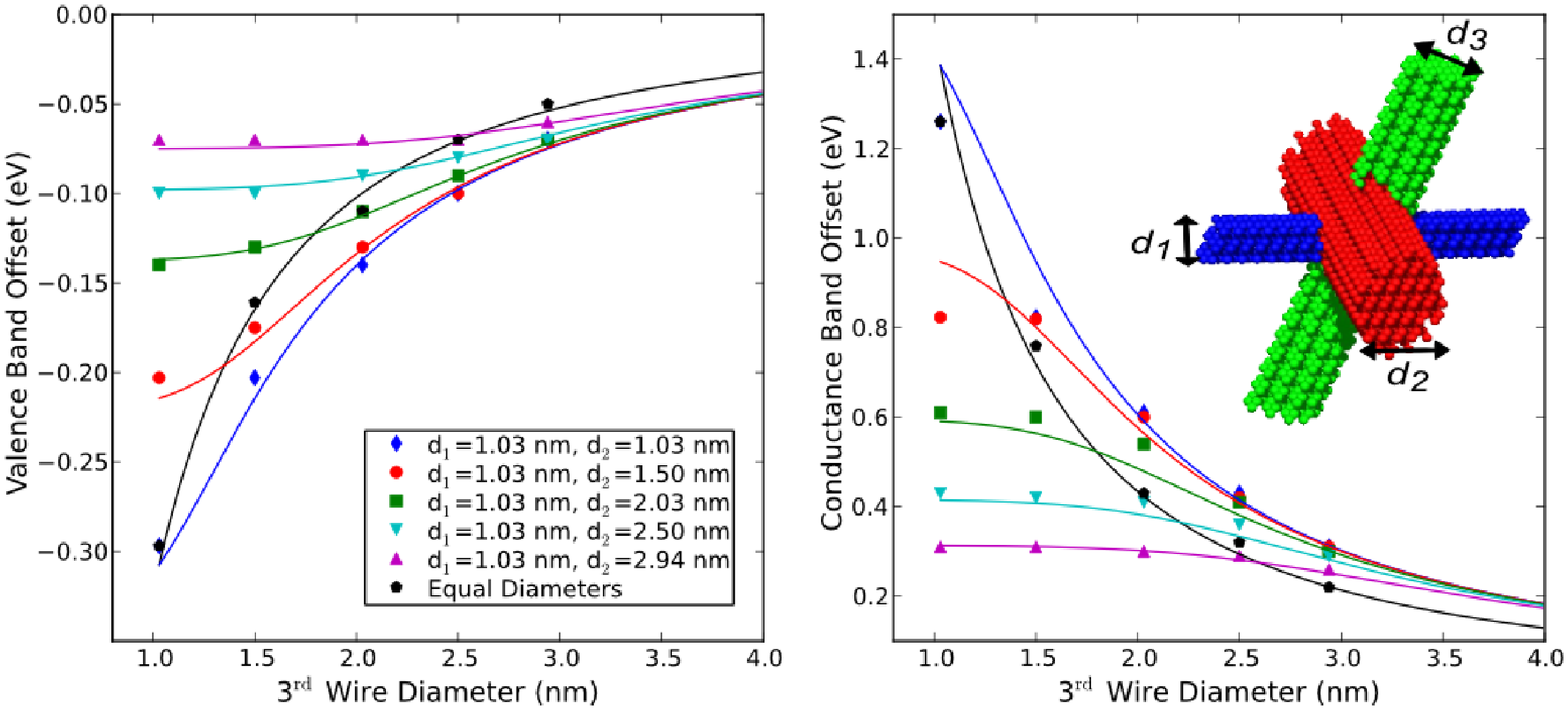}
\caption{\label{vcbo} Valence and conduction band offsets for
$\langle110\rangle$ three-wire crossings with respect to bulk Si
band edges. The calculated values are shown with the markers
whereas the estimation values based on the single nanowire
parameters are plotted with the curves. }
\end{Sfigure}

%



\begin{thebibliography}{43}%
\makeatletter
\providecommand \@ifxundefined [1]{%
 \@ifx{#1\undefined}
}%
\providecommand \@ifnum [1]{%
 \ifnum #1\expandafter \@firstoftwo
 \else \expandafter \@secondoftwo
 \fi
}%
\providecommand \@ifx [1]{%
 \ifx #1\expandafter \@firstoftwo
 \else \expandafter \@secondoftwo
 \fi
}%
\providecommand \natexlab [1]{#1}%
\providecommand \enquote  [1]{``#1''}%
\providecommand \bibnamefont  [1]{#1}%
\providecommand \bibfnamefont [1]{#1}%
\providecommand \citenamefont [1]{#1}%
\providecommand \href@noop [0]{\@secondoftwo}%
\providecommand \href [0]{\begingroup \@sanitize@url \@href}%
\providecommand \@href[1]{\@@startlink{#1}\@@href}%
\providecommand \@@href[1]{\endgroup#1\@@endlink}%
\providecommand \@sanitize@url [0]{\catcode `\\12\catcode
`\$12\catcode
  `\&12\catcode `\#12\catcode `\^12\catcode `\_12\catcode `\%12\relax}%
\providecommand \@@startlink[1]{}%
\providecommand \@@endlink[0]{}%
\providecommand \url  [0]{\begingroup\@sanitize@url \@url }%
\providecommand \@url [1]{\endgroup\@href {#1}{\urlprefix }}%
\providecommand \urlprefix  [0]{URL }%
\providecommand \Eprint [0]{\href }%
\providecommand \doibase [0]{http://dx.doi.org/}%
\providecommand \selectlanguage [0]{\@gobble}%
\providecommand \bibinfo  [0]{\@secondoftwo}%
\providecommand \bibfield  [0]{\@secondoftwo}%
\providecommand \translation [1]{[#1]}%
\providecommand \BibitemOpen [0]{}%
\providecommand \bibitemStop [0]{}%
\providecommand \bibitemNoStop [0]{.\EOS\space}%
\providecommand \EOS [0]{\spacefactor3000\relax}%
\providecommand \BibitemShut  [1]{\csname bibitem#1\endcsname}%
\let\auto@bib@innerbib\@empty
\bibitem [{\citenamefont {Lu}, \citenamefont {Xie},\ and\ \citenamefont
  {Lieber}(2008)}]{lu2008}%
  \BibitemOpen
  \bibfield  {author} {\bibinfo {author} {\bibfnamefont {W.}~\bibnamefont
  {Lu}}, \bibinfo {author} {\bibfnamefont {P.}~\bibnamefont {Xie}}, \ and\
  \bibinfo {author} {\bibfnamefont {C.~M.}\ \bibnamefont {Lieber}},\ }\href
  {\doibase 10.1109/TED.2008.2005158} {\bibfield  {journal} {\bibinfo
  {journal} {IEEE Trans. Electron Devices}\ }\textbf {\bibinfo {volume} {55}},\
  \bibinfo {pages} {2859} (\bibinfo {year} {2008})}\BibitemShut {NoStop}%
\bibitem [{\citenamefont {Morales}\ and\ \citenamefont
  {Lieber}(1998)}]{morales1998}%
  \BibitemOpen
  \bibfield  {author} {\bibinfo {author} {\bibfnamefont {A.~M.}\ \bibnamefont
  {Morales}}\ and\ \bibinfo {author} {\bibfnamefont {C.~M.}\ \bibnamefont
  {Lieber}},\ }\href {\doibase 10.1126/science.279.5348.208} {\bibfield
  {journal} {\bibinfo  {journal} {Science}\ }\textbf {\bibinfo {volume}
  {279}},\ \bibinfo {pages} {208} (\bibinfo {year} {1998})}\BibitemShut
  {NoStop}%
\bibitem [{\citenamefont {Holmes}\ \emph {et~al.}(2000)\citenamefont {Holmes},
  \citenamefont {Johnston}, \citenamefont {Doty},\ and\ \citenamefont
  {Korgel}}]{holmes2000}%
  \BibitemOpen
  \bibfield  {author} {\bibinfo {author} {\bibfnamefont {J.~D.}\ \bibnamefont
  {Holmes}}, \bibinfo {author} {\bibfnamefont {K.~P.}\ \bibnamefont
  {Johnston}}, \bibinfo {author} {\bibfnamefont {R.~C.}\ \bibnamefont {Doty}},
  \ and\ \bibinfo {author} {\bibfnamefont {B.~A.}\ \bibnamefont {Korgel}},\
  }\href {\doibase 10.1126/science.287.5457.1471} {\bibfield  {journal}
  {\bibinfo  {journal} {Science}\ }\textbf {\bibinfo {volume} {287}},\ \bibinfo
  {pages} {1471} (\bibinfo {year} {2000})}\BibitemShut {NoStop}%
\bibitem [{\citenamefont {Cui}\ \emph {et~al.}(2001)\citenamefont {Cui},
  \citenamefont {Lauhon}, \citenamefont {Gudiksen}, \citenamefont {Wang},\ and\
  \citenamefont {Lieber}}]{cui2001wire}%
  \BibitemOpen
  \bibfield  {author} {\bibinfo {author} {\bibfnamefont {Y.}~\bibnamefont
  {Cui}}, \bibinfo {author} {\bibfnamefont {L.~J.}\ \bibnamefont {Lauhon}},
  \bibinfo {author} {\bibfnamefont {M.~S.}\ \bibnamefont {Gudiksen}}, \bibinfo
  {author} {\bibfnamefont {J.}~\bibnamefont {Wang}}, \ and\ \bibinfo {author}
  {\bibfnamefont {C.~M.}\ \bibnamefont {Lieber}},\ }\href {\doibase
  10.1063/1.1363692} {\bibfield  {journal} {\bibinfo  {journal} {Appl. Phys.
  Lett.}\ }\textbf {\bibinfo {volume} {78}},\ \bibinfo {pages} {2214} (\bibinfo
  {year} {2001})}\BibitemShut {NoStop}%
\bibitem [{\citenamefont {Ma}\ \emph {et~al.}(2003)\citenamefont {Ma},
  \citenamefont {Lee}, \citenamefont {Au}, \citenamefont {Tong},\ and\
  \citenamefont {Lee}}]{ma2003}%
  \BibitemOpen
  \bibfield  {author} {\bibinfo {author} {\bibfnamefont {D.~D.~D.}\
  \bibnamefont {Ma}}, \bibinfo {author} {\bibfnamefont {C.~S.}\ \bibnamefont
  {Lee}}, \bibinfo {author} {\bibfnamefont {F.~C.~K.}\ \bibnamefont {Au}},
  \bibinfo {author} {\bibfnamefont {S.~Y.}\ \bibnamefont {Tong}}, \ and\
  \bibinfo {author} {\bibfnamefont {S.~T.}\ \bibnamefont {Lee}},\ }\href
  {\doibase 10.1126/science.1080313} {\bibfield  {journal} {\bibinfo  {journal}
  {Science}\ }\textbf {\bibinfo {volume} {299}},\ \bibinfo {pages} {1874}
  (\bibinfo {year} {2003})}\BibitemShut {NoStop}%
\bibitem [{\citenamefont {Wu}\ \emph {et~al.}(2004)\citenamefont {Wu},
  \citenamefont {Cui}, \citenamefont {Huynh}, \citenamefont {Barrelet},
  \citenamefont {Bell},\ and\ \citenamefont {Lieber}}]{wu2004}%
  \BibitemOpen
  \bibfield  {author} {\bibinfo {author} {\bibfnamefont {Y.}~\bibnamefont
  {Wu}}, \bibinfo {author} {\bibfnamefont {Y.}~\bibnamefont {Cui}}, \bibinfo
  {author} {\bibfnamefont {L.}~\bibnamefont {Huynh}}, \bibinfo {author}
  {\bibfnamefont {C.~J.}\ \bibnamefont {Barrelet}}, \bibinfo {author}
  {\bibfnamefont {D.~C.}\ \bibnamefont {Bell}}, \ and\ \bibinfo {author}
  {\bibfnamefont {C.~M.}\ \bibnamefont {Lieber}},\ }\href {\doibase
  10.1021/nl035162i} {\bibfield  {journal} {\bibinfo  {journal} {Nano Lett.}\
  }\textbf {\bibinfo {volume} {4}},\ \bibinfo {pages} {433} (\bibinfo {year}
  {2004})}\BibitemShut {NoStop}%
\bibitem [{\citenamefont {Manna}\ \emph {et~al.}(2003)\citenamefont {Manna},
  \citenamefont {Milliron}, \citenamefont {Meisel}, \citenamefont {Scher},\
  and\ \citenamefont {Alivisatos}}]{manna2003}%
  \BibitemOpen
  \bibfield  {author} {\bibinfo {author} {\bibfnamefont {L.}~\bibnamefont
  {Manna}}, \bibinfo {author} {\bibfnamefont {D.}~\bibnamefont {Milliron}},
  \bibinfo {author} {\bibfnamefont {A.}~\bibnamefont {Meisel}}, \bibinfo
  {author} {\bibfnamefont {E.}~\bibnamefont {Scher}}, \ and\ \bibinfo {author}
  {\bibfnamefont {A.}~\bibnamefont {Alivisatos}},\ }\href
  {http://www.nature.com/nmat/journal/v2/n6/abs/nmat902.html} {\bibfield
  {journal} {\bibinfo  {journal} {Nature Mater.}\ }\textbf {\bibinfo {volume}
  {2}},\ \bibinfo {pages} {382} (\bibinfo {year} {2003})}\BibitemShut {NoStop}%
\bibitem [{\citenamefont {Wang}\ \emph {et~al.}(2004)\citenamefont {Wang},
  \citenamefont {Qian}, \citenamefont {Yang}, \citenamefont {Zhong},\ and\
  \citenamefont {Lieber}}]{wang2004}%
  \BibitemOpen
  \bibfield  {author} {\bibinfo {author} {\bibfnamefont {D.}~\bibnamefont
  {Wang}}, \bibinfo {author} {\bibfnamefont {F.}~\bibnamefont {Qian}}, \bibinfo
  {author} {\bibfnamefont {C.}~\bibnamefont {Yang}}, \bibinfo {author}
  {\bibfnamefont {Z.}~\bibnamefont {Zhong}}, \ and\ \bibinfo {author}
  {\bibfnamefont {C.~M.}\ \bibnamefont {Lieber}},\ }\href {\doibase
  10.1021/nl049728u} {\bibfield  {journal} {\bibinfo  {journal} {Nano Lett.}\
  }\textbf {\bibinfo {volume} {4}},\ \bibinfo {pages} {871} (\bibinfo {year}
  {2004})}\BibitemShut {NoStop}%
\bibitem [{\citenamefont {Jun}\ and\ \citenamefont {Jacobson}(2010)}]{jun2010}%
  \BibitemOpen
  \bibfield  {author} {\bibinfo {author} {\bibfnamefont {K.}~\bibnamefont
  {Jun}}\ and\ \bibinfo {author} {\bibfnamefont {J.~M.}\ \bibnamefont
  {Jacobson}},\ }\href {\doibase 10.1021/nl100662z} {\bibfield  {journal}
  {\bibinfo  {journal} {Nano Lett.}\ }\textbf {\bibinfo {volume} {10}},\
  \bibinfo {pages} {2777} (\bibinfo {year} {2010})}\BibitemShut {NoStop}%
\bibitem [{\citenamefont {Jiang}\ \emph {et~al.}(2011)\citenamefont {Jiang},
  \citenamefont {Tian}, \citenamefont {Xiang}, \citenamefont {Qian},
  \citenamefont {Zheng}, \citenamefont {Wang}, \citenamefont {Mai},\ and\
  \citenamefont {Lieber}}]{jiang2011}%
  \BibitemOpen
  \bibfield  {author} {\bibinfo {author} {\bibfnamefont {X.}~\bibnamefont
  {Jiang}}, \bibinfo {author} {\bibfnamefont {B.}~\bibnamefont {Tian}},
  \bibinfo {author} {\bibfnamefont {J.}~\bibnamefont {Xiang}}, \bibinfo
  {author} {\bibfnamefont {F.}~\bibnamefont {Qian}}, \bibinfo {author}
  {\bibfnamefont {G.}~\bibnamefont {Zheng}}, \bibinfo {author} {\bibfnamefont
  {H.}~\bibnamefont {Wang}}, \bibinfo {author} {\bibfnamefont {L.}~\bibnamefont
  {Mai}}, \ and\ \bibinfo {author} {\bibfnamefont {C.}~\bibnamefont {Lieber}},\
  }\href {http://www.pnas.org/content/108/30/12212.short} {\bibfield  {journal}
  {\bibinfo  {journal} {Proc. Natl. Acad. Sci. U.S.A.}\ }\textbf {\bibinfo
  {volume} {108}},\ \bibinfo {pages} {12212} (\bibinfo {year}
  {2011})}\BibitemShut {NoStop}%
\bibitem [{\citenamefont {Dick}\ \emph {et~al.}(2004)\citenamefont {Dick},
  \citenamefont {Deppert}, \citenamefont {Larsson}, \citenamefont
  {M{\aa}rtensson}, \citenamefont {Seifert}, \citenamefont {Wallenberg},\ and\
  \citenamefont {Samuelson}}]{dick2004}%
  \BibitemOpen
  \bibfield  {author} {\bibinfo {author} {\bibfnamefont {K.}~\bibnamefont
  {Dick}}, \bibinfo {author} {\bibfnamefont {K.}~\bibnamefont {Deppert}},
  \bibinfo {author} {\bibfnamefont {M.}~\bibnamefont {Larsson}}, \bibinfo
  {author} {\bibfnamefont {T.}~\bibnamefont {M{\aa}rtensson}}, \bibinfo
  {author} {\bibfnamefont {W.}~\bibnamefont {Seifert}}, \bibinfo {author}
  {\bibfnamefont {L.}~\bibnamefont {Wallenberg}}, \ and\ \bibinfo {author}
  {\bibfnamefont {L.}~\bibnamefont {Samuelson}},\ }\href
  {http://www.nature.com/nmat/journal/v3/n6/abs/nmat1133.html} {\bibfield
  {journal} {\bibinfo  {journal} {Nature Mater.}\ }\textbf {\bibinfo {volume}
  {3}},\ \bibinfo {pages} {380} (\bibinfo {year} {2004})}\BibitemShut {NoStop}%
\bibitem [{\citenamefont {Zhou}\ \emph {et~al.}(2005)\citenamefont {Zhou},
  \citenamefont {Ding}, \citenamefont {Deng}, \citenamefont {Gong},
  \citenamefont {Xu},\ and\ \citenamefont {Wang}}]{zhou2005wo}%
  \BibitemOpen
  \bibfield  {author} {\bibinfo {author} {\bibfnamefont {J.}~\bibnamefont
  {Zhou}}, \bibinfo {author} {\bibfnamefont {Y.}~\bibnamefont {Ding}}, \bibinfo
  {author} {\bibfnamefont {S.}~\bibnamefont {Deng}}, \bibinfo {author}
  {\bibfnamefont {L.}~\bibnamefont {Gong}}, \bibinfo {author} {\bibfnamefont
  {N.}~\bibnamefont {Xu}}, \ and\ \bibinfo {author} {\bibfnamefont
  {Z.}~\bibnamefont {Wang}},\ }\href {\doibase 10.1002/adma.200500885}
  {\bibfield  {journal} {\bibinfo  {journal} {Adv. Mater.}\ }\textbf {\bibinfo
  {volume} {17}},\ \bibinfo {pages} {2107} (\bibinfo {year}
  {2005})}\BibitemShut {NoStop}%
\bibitem [{\citenamefont {Dick}\ \emph {et~al.}(2006)\citenamefont {Dick},
  \citenamefont {Deppert}, \citenamefont {Karlsson}, \citenamefont {Seifert},
  \citenamefont {Wallenberg},\ and\ \citenamefont {Samuelson}}]{dick2006}%
  \BibitemOpen
  \bibfield  {author} {\bibinfo {author} {\bibfnamefont {K.~A.}\ \bibnamefont
  {Dick}}, \bibinfo {author} {\bibfnamefont {K.}~\bibnamefont {Deppert}},
  \bibinfo {author} {\bibfnamefont {L.~S.}\ \bibnamefont {Karlsson}}, \bibinfo
  {author} {\bibfnamefont {W.}~\bibnamefont {Seifert}}, \bibinfo {author}
  {\bibfnamefont {L.~R.}\ \bibnamefont {Wallenberg}}, \ and\ \bibinfo {author}
  {\bibfnamefont {L.}~\bibnamefont {Samuelson}},\ }\href {\doibase
  10.1021/nl062035o} {\bibfield  {journal} {\bibinfo  {journal} {Nano Lett.}\
  }\textbf {\bibinfo {volume} {6}},\ \bibinfo {pages} {2842} (\bibinfo {year}
  {2006})}\BibitemShut {NoStop}%
\bibitem [{\citenamefont {Totaro}, \citenamefont {Bruschi},\ and\ \citenamefont
  {Pennelli}(2012)}]{totaro2012}%
  \BibitemOpen
  \bibfield  {author} {\bibinfo {author} {\bibfnamefont {M.}~\bibnamefont
  {Totaro}}, \bibinfo {author} {\bibfnamefont {P.}~\bibnamefont {Bruschi}}, \
  and\ \bibinfo {author} {\bibfnamefont {G.}~\bibnamefont {Pennelli}},\ }\href
  {\doibase http://dx.doi.org/10.1016/j.mee.2012.04.007} {\bibfield  {journal}
  {\bibinfo  {journal} {Microelectron. Eng.}\ }\textbf {\bibinfo {volume}
  {97}},\ \bibinfo {pages} {157 } (\bibinfo {year} {2012})}\BibitemShut
  {NoStop}%
\bibitem [{\citenamefont {Serre}\ \emph {et~al.}(2013)\citenamefont {Serre},
  \citenamefont {Ternon}, \citenamefont {Stambouli}, \citenamefont {Periwal},\
  and\ \citenamefont {Baron}}]{serre2013}%
  \BibitemOpen
  \bibfield  {author} {\bibinfo {author} {\bibfnamefont {P.}~\bibnamefont
  {Serre}}, \bibinfo {author} {\bibfnamefont {C.}~\bibnamefont {Ternon}},
  \bibinfo {author} {\bibfnamefont {V.}~\bibnamefont {Stambouli}}, \bibinfo
  {author} {\bibfnamefont {P.}~\bibnamefont {Periwal}}, \ and\ \bibinfo
  {author} {\bibfnamefont {T.}~\bibnamefont {Baron}},\ }\href {\doibase
  http://dx.doi.org/10.1016/j.snb.2013.03.022} {\bibfield  {journal} {\bibinfo
  {journal} {Sens. Actuators, B}\ }\textbf {\bibinfo {volume} {182}},\ \bibinfo
  {pages} {390 } (\bibinfo {year} {2013})}\BibitemShut {NoStop}%
\bibitem [{\citenamefont {Mulazimoglu}\ \emph {et~al.}(2013)\citenamefont
  {Mulazimoglu}, \citenamefont {Coskun}, \citenamefont {Gunoven}, \citenamefont
  {Butun}, \citenamefont {Ozbay}, \citenamefont {Turan},\ and\ \citenamefont
  {Unalan}}]{mulazimoglu2013}%
  \BibitemOpen
  \bibfield  {author} {\bibinfo {author} {\bibfnamefont {E.}~\bibnamefont
  {Mulazimoglu}}, \bibinfo {author} {\bibfnamefont {S.}~\bibnamefont {Coskun}},
  \bibinfo {author} {\bibfnamefont {M.}~\bibnamefont {Gunoven}}, \bibinfo
  {author} {\bibfnamefont {B.}~\bibnamefont {Butun}}, \bibinfo {author}
  {\bibfnamefont {E.}~\bibnamefont {Ozbay}}, \bibinfo {author} {\bibfnamefont
  {R.}~\bibnamefont {Turan}}, \ and\ \bibinfo {author} {\bibfnamefont {H.~E.}\
  \bibnamefont {Unalan}},\ }\href {\doibase 10.1063/1.4819387} {\bibfield
  {journal} {\bibinfo  {journal} {Appl. Phys. Lett.}\ }\textbf {\bibinfo
  {volume} {103}},\ \bibinfo {eid} {083114} (\bibinfo {year}
  {2013})}\BibitemShut {NoStop}%
\bibitem [{\citenamefont {Zhao}\ \emph {et~al.}(2004)\citenamefont {Zhao},
  \citenamefont {Wei}, \citenamefont {Yang},\ and\ \citenamefont
  {Chou}}]{zhao2004}%
  \BibitemOpen
  \bibfield  {author} {\bibinfo {author} {\bibfnamefont {X.}~\bibnamefont
  {Zhao}}, \bibinfo {author} {\bibfnamefont {C.~M.}\ \bibnamefont {Wei}},
  \bibinfo {author} {\bibfnamefont {L.}~\bibnamefont {Yang}}, \ and\ \bibinfo
  {author} {\bibfnamefont {M.~Y.}\ \bibnamefont {Chou}},\ }\href {\doibase
  10.1103/PhysRevLett.92.236805} {\bibfield  {journal} {\bibinfo  {journal}
  {Phys. Rev. Lett.}\ }\textbf {\bibinfo {volume} {92}},\ \bibinfo {pages}
  {236805} (\bibinfo {year} {2004})}\BibitemShut {NoStop}%
\bibitem [{\citenamefont {Bruno}\ \emph {et~al.}(2007)\citenamefont {Bruno},
  \citenamefont {Palummo}, \citenamefont {Marini}, \citenamefont {Del~Sole},\
  and\ \citenamefont {Ossicini}}]{bruno2007}%
  \BibitemOpen
  \bibfield  {author} {\bibinfo {author} {\bibfnamefont {M.}~\bibnamefont
  {Bruno}}, \bibinfo {author} {\bibfnamefont {M.}~\bibnamefont {Palummo}},
  \bibinfo {author} {\bibfnamefont {A.}~\bibnamefont {Marini}}, \bibinfo
  {author} {\bibfnamefont {R.}~\bibnamefont {Del~Sole}}, \ and\ \bibinfo
  {author} {\bibfnamefont {S.}~\bibnamefont {Ossicini}},\ }\href {\doibase
  10.1103/PhysRevLett.98.036807} {\bibfield  {journal} {\bibinfo  {journal}
  {Phys. Rev. Lett.}\ }\textbf {\bibinfo {volume} {98}},\ \bibinfo {pages}
  {036807} (\bibinfo {year} {2007})}\BibitemShut {NoStop}%
\bibitem [{\citenamefont {Yan}, \citenamefont {Yang},\ and\ \citenamefont
  {Chou}(2007)}]{yan2007}%
  \BibitemOpen
  \bibfield  {author} {\bibinfo {author} {\bibfnamefont {J.-A.}\ \bibnamefont
  {Yan}}, \bibinfo {author} {\bibfnamefont {L.}~\bibnamefont {Yang}}, \ and\
  \bibinfo {author} {\bibfnamefont {M.~Y.}\ \bibnamefont {Chou}},\ }\href
  {\doibase 10.1103/PhysRevB.76.115319} {\bibfield  {journal} {\bibinfo
  {journal} {Phys. Rev. B}\ }\textbf {\bibinfo {volume} {76}},\ \bibinfo
  {pages} {115319} (\bibinfo {year} {2007})}\BibitemShut {NoStop}%
\bibitem [{\citenamefont {Huang}\ \emph {et~al.}(2008)\citenamefont {Huang},
  \citenamefont {Lu}, \citenamefont {Yan}, \citenamefont {Chou}, \citenamefont
  {Wang},\ and\ \citenamefont {Ho}}]{huang2008}%
  \BibitemOpen
  \bibfield  {author} {\bibinfo {author} {\bibfnamefont {L.}~\bibnamefont
  {Huang}}, \bibinfo {author} {\bibfnamefont {N.}~\bibnamefont {Lu}}, \bibinfo
  {author} {\bibfnamefont {J.-A.}\ \bibnamefont {Yan}}, \bibinfo {author}
  {\bibfnamefont {M.~Y.}\ \bibnamefont {Chou}}, \bibinfo {author}
  {\bibfnamefont {C.-Z.}\ \bibnamefont {Wang}}, \ and\ \bibinfo {author}
  {\bibfnamefont {K.-M.}\ \bibnamefont {Ho}},\ }\href {\doibase
  10.1021/jp802591v} {\bibfield  {journal} {\bibinfo  {journal} {J. Phys. Chem.
  C}\ }\textbf {\bibinfo {volume} {112}},\ \bibinfo {pages} {15680} (\bibinfo
  {year} {2008})}\BibitemShut {NoStop}%
\bibitem [{\citenamefont {Rurali}\ \emph {et~al.}(2007)\citenamefont {Rurali},
  \citenamefont {Aradi}, \citenamefont {Frauenheim},\ and\ \citenamefont
  {Gali}}]{rurali2007}%
  \BibitemOpen
  \bibfield  {author} {\bibinfo {author} {\bibfnamefont {R.}~\bibnamefont
  {Rurali}}, \bibinfo {author} {\bibfnamefont {B.}~\bibnamefont {Aradi}},
  \bibinfo {author} {\bibfnamefont {T.}~\bibnamefont {Frauenheim}}, \ and\
  \bibinfo {author} {\bibfnamefont {A.}~\bibnamefont {Gali}},\ }\href {\doibase
  10.1103/PhysRevB.76.113303} {\bibfield  {journal} {\bibinfo  {journal} {Phys.
  Rev. B}\ }\textbf {\bibinfo {volume} {76}},\ \bibinfo {pages} {113303}
  (\bibinfo {year} {2007})}\BibitemShut {NoStop}%
\bibitem [{\citenamefont {Ng}\ \emph {et~al.}(2007)\citenamefont {Ng},
  \citenamefont {Zhou}, \citenamefont {Yang}, \citenamefont {Sim},
  \citenamefont {Tan},\ and\ \citenamefont {Wu}}]{ng2007}%
  \BibitemOpen
  \bibfield  {author} {\bibinfo {author} {\bibfnamefont {M.-F.}\ \bibnamefont
  {Ng}}, \bibinfo {author} {\bibfnamefont {L.}~\bibnamefont {Zhou}}, \bibinfo
  {author} {\bibfnamefont {S.-W.}\ \bibnamefont {Yang}}, \bibinfo {author}
  {\bibfnamefont {L.~Y.}\ \bibnamefont {Sim}}, \bibinfo {author} {\bibfnamefont
  {V.~B.~C.}\ \bibnamefont {Tan}}, \ and\ \bibinfo {author} {\bibfnamefont
  {P.}~\bibnamefont {Wu}},\ }\href {\doibase 10.1103/PhysRevB.76.155435}
  {\bibfield  {journal} {\bibinfo  {journal} {Phys. Rev. B}\ }\textbf {\bibinfo
  {volume} {76}},\ \bibinfo {pages} {155435} (\bibinfo {year}
  {2007})}\BibitemShut {NoStop}%
\bibitem [{\citenamefont {Scheel}, \citenamefont {Reich},\ and\ \citenamefont
  {Thomsen}(2005)}]{scheel2005}%
  \BibitemOpen
  \bibfield  {author} {\bibinfo {author} {\bibfnamefont {H.}~\bibnamefont
  {Scheel}}, \bibinfo {author} {\bibfnamefont {S.}~\bibnamefont {Reich}}, \
  and\ \bibinfo {author} {\bibfnamefont {C.}~\bibnamefont {Thomsen}},\ }\href
  {\doibase 10.1002/pssb.200541133} {\bibfield  {journal} {\bibinfo  {journal}
  {Phys. Status Solidi B}\ }\textbf {\bibinfo {volume} {242}},\ \bibinfo
  {pages} {2474} (\bibinfo {year} {2005})}\BibitemShut {NoStop}%
\bibitem [{\citenamefont {Niquet}\ \emph {et~al.}(2006)\citenamefont {Niquet},
  \citenamefont {Lherbier}, \citenamefont {Quang}, \citenamefont
  {Fern\'andez-Serra}, \citenamefont {Blase},\ and\ \citenamefont
  {Delerue}}]{niquet2006}%
  \BibitemOpen
  \bibfield  {author} {\bibinfo {author} {\bibfnamefont {Y.~M.}\ \bibnamefont
  {Niquet}}, \bibinfo {author} {\bibfnamefont {A.}~\bibnamefont {Lherbier}},
  \bibinfo {author} {\bibfnamefont {N.~H.}\ \bibnamefont {Quang}}, \bibinfo
  {author} {\bibfnamefont {M.~V.}\ \bibnamefont {Fern\'andez-Serra}}, \bibinfo
  {author} {\bibfnamefont {X.}~\bibnamefont {Blase}}, \ and\ \bibinfo {author}
  {\bibfnamefont {C.}~\bibnamefont {Delerue}},\ }\href {\doibase
  10.1103/PhysRevB.73.165319} {\bibfield  {journal} {\bibinfo  {journal} {Phys.
  Rev. B}\ }\textbf {\bibinfo {volume} {73}},\ \bibinfo {pages} {165319}
  (\bibinfo {year} {2006})}\BibitemShut {NoStop}%
\bibitem [{\citenamefont {Kim}\ and\ \citenamefont
  {Fischetti}(2011)}]{kim2011}%
  \BibitemOpen
  \bibfield  {author} {\bibinfo {author} {\bibfnamefont {J.}~\bibnamefont
  {Kim}}\ and\ \bibinfo {author} {\bibfnamefont {M.~V.}\ \bibnamefont
  {Fischetti}},\ }\href {\doibase 10.1063/1.3615942} {\bibfield  {journal}
  {\bibinfo  {journal} {J. Appl. Phys.}\ }\textbf {\bibinfo {volume} {110}},\
  \bibinfo {eid} {033716} (\bibinfo {year} {2011})}\BibitemShut {NoStop}%
\bibitem [{\citenamefont {Rurali}(2010)}]{ruralirev}%
  \BibitemOpen
  \bibfield  {author} {\bibinfo {author} {\bibfnamefont {R.}~\bibnamefont
  {Rurali}},\ }\href {\doibase 10.1103/RevModPhys.82.427} {\bibfield  {journal}
  {\bibinfo  {journal} {Rev. Mod. Phys.}\ }\textbf {\bibinfo {volume} {82}},\
  \bibinfo {pages} {427} (\bibinfo {year} {2010})}\BibitemShut {NoStop}%
\bibitem [{\citenamefont {Menon}\ \emph {et~al.}(2007)\citenamefont {Menon},
  \citenamefont {Richter}, \citenamefont {Lee},\ and\ \citenamefont
  {Raghavan}}]{menon2007}%
  \BibitemOpen
  \bibfield  {author} {\bibinfo {author} {\bibfnamefont {M.}~\bibnamefont
  {Menon}}, \bibinfo {author} {\bibfnamefont {E.}~\bibnamefont {Richter}},
  \bibinfo {author} {\bibfnamefont {I.}~\bibnamefont {Lee}}, \ and\ \bibinfo
  {author} {\bibfnamefont {P.}~\bibnamefont {Raghavan}},\ }\href {\doibase
  doi:10.1166/jctn.2007.008} {\bibfield  {journal} {\bibinfo  {journal} {J.
  Comput. Theor. Nanosci.}\ }\textbf {\bibinfo {volume} {4}},\ \bibinfo {pages}
  {252} (\bibinfo {year} {2007})}\BibitemShut {NoStop}%
\bibitem [{\citenamefont {Avramov}\ \emph {et~al.}(2007)\citenamefont
  {Avramov}, \citenamefont {Chernozatonskii}, \citenamefont {Sorokin},\ and\
  \citenamefont {Gordon}}]{avramov2007nano}%
  \BibitemOpen
  \bibfield  {author} {\bibinfo {author} {\bibfnamefont {P.~V.}\ \bibnamefont
  {Avramov}}, \bibinfo {author} {\bibfnamefont {L.~A.}\ \bibnamefont
  {Chernozatonskii}}, \bibinfo {author} {\bibfnamefont {P.~B.}\ \bibnamefont
  {Sorokin}}, \ and\ \bibinfo {author} {\bibfnamefont {M.~S.}\ \bibnamefont
  {Gordon}},\ }\href {\doibase 10.1021/nl070973y} {\bibfield  {journal}
  {\bibinfo  {journal} {Nano Lett.}\ }\textbf {\bibinfo {volume} {7}},\
  \bibinfo {pages} {2063} (\bibinfo {year} {2007})}\BibitemShut {NoStop}%
\bibitem [{\citenamefont {Bester}(2009)}]{bester}%
  \BibitemOpen
  \bibfield  {author} {\bibinfo {author} {\bibfnamefont {G.}~\bibnamefont
  {Bester}},\ }\href {http://stacks.iop.org/0953-8984/21/i=2/a=023202}
  {\bibfield  {journal} {\bibinfo  {journal} {J. Phys.: Condens. Matter}\
  }\textbf {\bibinfo {volume} {21}},\ \bibinfo {pages} {023202} (\bibinfo
  {year} {2009})}\BibitemShut {NoStop}%
\bibitem [{\citenamefont {Wang}, \citenamefont {Franceschetti},\ and\
  \citenamefont {Zunger}(1997)}]{lcbb1997}%
  \BibitemOpen
  \bibfield  {author} {\bibinfo {author} {\bibfnamefont {L.-W.}\ \bibnamefont
  {Wang}}, \bibinfo {author} {\bibfnamefont {A.}~\bibnamefont {Franceschetti}},
  \ and\ \bibinfo {author} {\bibfnamefont {A.}~\bibnamefont {Zunger}},\ }\href
  {\doibase 10.1103/PhysRevLett.78.2819} {\bibfield  {journal} {\bibinfo
  {journal} {Phys. Rev. Lett.}\ }\textbf {\bibinfo {volume} {78}},\ \bibinfo
  {pages} {2819} (\bibinfo {year} {1997})}\BibitemShut {NoStop}%
\bibitem [{\citenamefont {Wang}\ and\ \citenamefont {Zunger}(1999)}]{lcbb1999}%
  \BibitemOpen
  \bibfield  {author} {\bibinfo {author} {\bibfnamefont {L.-W.}\ \bibnamefont
  {Wang}}\ and\ \bibinfo {author} {\bibfnamefont {A.}~\bibnamefont {Zunger}},\
  }\href {\doibase 10.1103/PhysRevB.59.15806} {\bibfield  {journal} {\bibinfo
  {journal} {Phys. Rev. B}\ }\textbf {\bibinfo {volume} {59}},\ \bibinfo
  {pages} {15806} (\bibinfo {year} {1999})}\BibitemShut {NoStop}%
\bibitem [{\citenamefont {Bulutay}(2007)}]{bulutay2007}%
  \BibitemOpen
  \bibfield  {author} {\bibinfo {author} {\bibfnamefont {C.}~\bibnamefont
  {Bulutay}},\ }\href {\doibase 10.1103/PhysRevB.76.205321} {\bibfield
  {journal} {\bibinfo  {journal} {Phys. Rev. B}\ }\textbf {\bibinfo {volume}
  {76}},\ \bibinfo {pages} {205321} (\bibinfo {year} {2007})}\BibitemShut
  {NoStop}%
\bibitem [{sup()}]{suppl_matl}%
  \BibitemOpen
  \href@noop {} {}\bibinfo {note} {See supplementary material (appended to this manuscript)
  for the details of the implementation of LCBB method,
  pseudopotential form factors, generalized mean, and estimation expression for
  valence and conduction band offsets.}\BibitemShut {Stop}%
\bibitem [{\citenamefont {Y{\i}ld{\i}r{\i}m}\ and\ \citenamefont
  {Bulutay}(2008)}]{yildirim2008}%
  \BibitemOpen
  \bibfield  {author} {\bibinfo {author} {\bibfnamefont {H.}~\bibnamefont
  {Y{\i}ld{\i}r{\i}m}}\ and\ \bibinfo {author} {\bibfnamefont {C.}~\bibnamefont
  {Bulutay}},\ }\href {\doibase 10.1103/PhysRevB.78.115307} {\bibfield
  {journal} {\bibinfo  {journal} {Phys. Rev. B}\ }\textbf {\bibinfo {volume}
  {78}},\ \bibinfo {pages} {115307} (\bibinfo {year} {2008})}\BibitemShut
  {NoStop}%
\bibitem [{\citenamefont {Imakita}\ \emph {et~al.}(2009)\citenamefont
  {Imakita}, \citenamefont {Ito}, \citenamefont {Fujii},\ and\ \citenamefont
  {Hayashi}}]{imakita2009}%
  \BibitemOpen
  \bibfield  {author} {\bibinfo {author} {\bibfnamefont {K.}~\bibnamefont
  {Imakita}}, \bibinfo {author} {\bibfnamefont {M.}~\bibnamefont {Ito}},
  \bibinfo {author} {\bibfnamefont {M.}~\bibnamefont {Fujii}}, \ and\ \bibinfo
  {author} {\bibfnamefont {S.}~\bibnamefont {Hayashi}},\ }\href {\doibase
  10.1063/1.3125446} {\bibfield  {journal} {\bibinfo  {journal} {J. Appl.
  Phys.}\ }\textbf {\bibinfo {volume} {105}},\ \bibinfo {eid} {093531}
  (\bibinfo {year} {2009})}\BibitemShut {NoStop}%
\bibitem [{\citenamefont {Bulutay}, \citenamefont {Kulakci},\ and\
  \citenamefont {Turan}(2010)}]{bulutay2010}%
  \BibitemOpen
  \bibfield  {author} {\bibinfo {author} {\bibfnamefont {C.}~\bibnamefont
  {Bulutay}}, \bibinfo {author} {\bibfnamefont {M.}~\bibnamefont {Kulakci}}, \
  and\ \bibinfo {author} {\bibfnamefont {R.}~\bibnamefont {Turan}},\ }\href
  {\doibase 10.1103/PhysRevB.81.125333} {\bibfield  {journal} {\bibinfo
  {journal} {Phys. Rev. B}\ }\textbf {\bibinfo {volume} {81}},\ \bibinfo
  {pages} {125333} (\bibinfo {year} {2010})}\BibitemShut {NoStop}%
\bibitem [{\citenamefont {Delerue}, \citenamefont {Allan},\ and\ \citenamefont
  {Lannoo}(1993)}]{delerue1993}%
  \BibitemOpen
  \bibfield  {author} {\bibinfo {author} {\bibfnamefont {C.}~\bibnamefont
  {Delerue}}, \bibinfo {author} {\bibfnamefont {G.}~\bibnamefont {Allan}}, \
  and\ \bibinfo {author} {\bibfnamefont {M.}~\bibnamefont {Lannoo}},\ }\href
  {\doibase 10.1103/PhysRevB.48.11024} {\bibfield  {journal} {\bibinfo
  {journal} {Phys. Rev. B}\ }\textbf {\bibinfo {volume} {48}},\ \bibinfo
  {pages} {11024} (\bibinfo {year} {1993})}\BibitemShut {NoStop}%
\bibitem [{\citenamefont {Brus}(1983)}]{brus83}%
  \BibitemOpen
  \bibfield  {author} {\bibinfo {author} {\bibfnamefont {L.~E.}\ \bibnamefont
  {Brus}},\ }\href {\doibase 10.1063/1.445676} {\bibfield  {journal} {\bibinfo
  {journal} {J. Chem. Phys.}\ }\textbf {\bibinfo {volume} {79}},\ \bibinfo
  {pages} {5566} (\bibinfo {year} {1983})}\BibitemShut {NoStop}%
\bibitem [{\citenamefont {Takagahara}\ and\ \citenamefont
  {Takeda}(1992)}]{takagahara92}%
  \BibitemOpen
  \bibfield  {author} {\bibinfo {author} {\bibfnamefont {T.}~\bibnamefont
  {Takagahara}}\ and\ \bibinfo {author} {\bibfnamefont {K.}~\bibnamefont
  {Takeda}},\ }\href {\doibase 10.1103/PhysRevB.46.15578} {\bibfield  {journal}
  {\bibinfo  {journal} {Phys. Rev. B}\ }\textbf {\bibinfo {volume} {46}},\
  \bibinfo {pages} {15578} (\bibinfo {year} {1992})}\BibitemShut {NoStop}%
\bibitem [{edg()}]{edgenote}%
  \BibitemOpen
  \href@noop {} {}\bibinfo {note} {This dual character of the valence band edge
  exponents is preserved, if we try the form $C_1 d^{-1}+C_2 d^{-2}$ as
  suggested in Ref. \onlinecite{ossicini1997}.}\BibitemShut {Stop}%
\bibitem [{\citenamefont {Bondi}, \citenamefont {Lee},\ and\ \citenamefont
  {Hwang}(2011)}]{bondi2011}%
  \BibitemOpen
  \bibfield  {author} {\bibinfo {author} {\bibfnamefont {R.~J.}\ \bibnamefont
  {Bondi}}, \bibinfo {author} {\bibfnamefont {S.}~\bibnamefont {Lee}}, \ and\
  \bibinfo {author} {\bibfnamefont {G.~S.}\ \bibnamefont {Hwang}},\ }\href
  {\doibase 10.1021/nn102232u} {\bibfield  {journal} {\bibinfo  {journal} {ACS
  Nano}\ }\textbf {\bibinfo {volume} {5}},\ \bibinfo {pages} {1713} (\bibinfo
  {year} {2011})}\BibitemShut {NoStop}%
\bibitem [{\citenamefont {Schmidt}, \citenamefont {Senz},\ and\ \citenamefont
  {G\"{o}sele}(2005)}]{schmidt2005}%
  \BibitemOpen
  \bibfield  {author} {\bibinfo {author} {\bibfnamefont {V.}~\bibnamefont
  {Schmidt}}, \bibinfo {author} {\bibfnamefont {S.}~\bibnamefont {Senz}}, \
  and\ \bibinfo {author} {\bibfnamefont {U.}~\bibnamefont {G\"{o}sele}},\
  }\href {\doibase 10.1021/nl050462g} {\bibfield  {journal} {\bibinfo
  {journal} {Nano Lett.}\ }\textbf {\bibinfo {volume} {5}},\ \bibinfo {pages}
  {931} (\bibinfo {year} {2005})}\BibitemShut {NoStop}%
\bibitem [{\citenamefont {Ossicini}\ \emph {et~al.}(1997)\citenamefont
  {Ossicini}, \citenamefont {Bertoni}, \citenamefont {Biagini}, \citenamefont
  {Lugli}, \citenamefont {Roma},\ and\ \citenamefont {Bisi}}]{ossicini1997}%
  \BibitemOpen
  \bibfield  {author} {\bibinfo {author} {\bibfnamefont {S.}~\bibnamefont
  {Ossicini}}, \bibinfo {author} {\bibfnamefont {C.}~\bibnamefont {Bertoni}},
  \bibinfo {author} {\bibfnamefont {M.}~\bibnamefont {Biagini}}, \bibinfo
  {author} {\bibfnamefont {A.}~\bibnamefont {Lugli}}, \bibinfo {author}
  {\bibfnamefont {G.}~\bibnamefont {Roma}}, \ and\ \bibinfo {author}
  {\bibfnamefont {O.}~\bibnamefont {Bisi}},\ }\href {\doibase
  10.1016/S0040-6090(96)09442-4} {\bibfield  {journal} {\bibinfo  {journal}
  {Thin Solid Films}\ }\textbf {\bibinfo {volume} {297}},\ \bibinfo {pages}
  {154} (\bibinfo {year} {1997})}\BibitemShut {NoStop}%
\end{thebibliography}

\begin{thebibliography}{9}%
\makeatletter
\providecommand \@ifxundefined [1]{%
 \@ifx{#1\undefined}
}%
\providecommand \@ifnum [1]{%
 \ifnum #1\expandafter \@firstoftwo
 \else \expandafter \@secondoftwo
 \fi
}%
\providecommand \@ifx [1]{%
 \ifx #1\expandafter \@firstoftwo
 \else \expandafter \@secondoftwo
 \fi
}%
\providecommand \natexlab [1]{#1}%
\providecommand \enquote  [1]{``#1''}%
\providecommand \bibnamefont  [1]{#1}%
\providecommand \bibfnamefont [1]{#1}%
\providecommand \citenamefont [1]{#1}%
\providecommand \href@noop [0]{\@secondoftwo}%
\providecommand \href [0]{\begingroup \@sanitize@url \@href}%
\providecommand \@href[1]{\@@startlink{#1}\@@href}%
\providecommand \@@href[1]{\endgroup#1\@@endlink}%
\providecommand \@sanitize@url [0]{\catcode `\\12\catcode
`\$12\catcode
  `\&12\catcode `\#12\catcode `\^12\catcode `\_12\catcode `\%12\relax}%
\providecommand \@@startlink[1]{}%
\providecommand \@@endlink[0]{}%
\providecommand \url  [0]{\begingroup\@sanitize@url \@url }%
\providecommand \@url [1]{\endgroup\@href {#1}{\urlprefix }}%
\providecommand \urlprefix  [0]{URL }%
\providecommand \Eprint [0]{\href }%
\providecommand \doibase [0]{http://dx.doi.org/}%
\providecommand \selectlanguage [0]{\@gobble}%
\providecommand \bibinfo  [0]{\@secondoftwo}%
\providecommand \bibfield  [0]{\@secondoftwo}%
\providecommand \translation [1]{[#1]}%
\providecommand \BibitemOpen [0]{}%
\providecommand \bibitemStop [0]{}%
\providecommand \bibitemNoStop [0]{.\EOS\space}%
\providecommand \EOS [0]{\spacefactor3000\relax}%
\providecommand \BibitemShut  [1]{\csname bibitem#1\endcsname}%
\let\auto@bib@innerbib\@empty
\bibitem [{\citenamefont {Ng}\ \emph {et~al.}(2007)\citenamefont {Ng},
  \citenamefont {Zhou}, \citenamefont {Yang}, \citenamefont {Sim},
  \citenamefont {Tan},\ and\ \citenamefont {Wu}}]{ng2007s}%
  \BibitemOpen
  \bibfield  {author} {\bibinfo {author} {\bibfnamefont {M.-F.}\ \bibnamefont
  {Ng}}, \bibinfo {author} {\bibfnamefont {L.}~\bibnamefont {Zhou}}, \bibinfo
  {author} {\bibfnamefont {S.-W.}\ \bibnamefont {Yang}}, \bibinfo {author}
  {\bibfnamefont {L.~Y.}\ \bibnamefont {Sim}}, \bibinfo {author} {\bibfnamefont
  {V.~B.~C.}\ \bibnamefont {Tan}}, \ and\ \bibinfo {author} {\bibfnamefont
  {P.}~\bibnamefont {Wu}},\ }\href {\doibase 10.1103/PhysRevB.76.155435}
  {\bibfield  {journal} {\bibinfo  {journal} {Phys. Rev. B}\ }\textbf {\bibinfo
  {volume} {76}},\ \bibinfo {pages} {155435} (\bibinfo {year}
  {2007})}\BibitemShut {NoStop}%
\bibitem [{\citenamefont {Niquet}\ \emph {et~al.}(2006)\citenamefont {Niquet},
  \citenamefont {Lherbier}, \citenamefont {Quang}, \citenamefont
  {Fern\'andez-Serra}, \citenamefont {Blase},\ and\ \citenamefont
  {Delerue}}]{niquet2006s}%
  \BibitemOpen
  \bibfield  {author} {\bibinfo {author} {\bibfnamefont {Y.~M.}\ \bibnamefont
  {Niquet}}, \bibinfo {author} {\bibfnamefont {A.}~\bibnamefont {Lherbier}},
  \bibinfo {author} {\bibfnamefont {N.~H.}\ \bibnamefont {Quang}}, \bibinfo
  {author} {\bibfnamefont {M.~V.}\ \bibnamefont {Fern\'andez-Serra}}, \bibinfo
  {author} {\bibfnamefont {X.}~\bibnamefont {Blase}}, \ and\ \bibinfo {author}
  {\bibfnamefont {C.}~\bibnamefont {Delerue}},\ }\href {\doibase
  10.1103/PhysRevB.73.165319} {\bibfield  {journal} {\bibinfo  {journal} {Phys.
  Rev. B}\ }\textbf {\bibinfo {volume} {73}},\ \bibinfo {pages} {165319}
  (\bibinfo {year} {2006})}\BibitemShut {NoStop}%
\bibitem [{\citenamefont {Friedel}, \citenamefont {Hybertsen},\ and\
  \citenamefont {Schl\"uter}(1989)}]{friedel1989s}%
  \BibitemOpen
  \bibfield  {author} {\bibinfo {author} {\bibfnamefont {P.}~\bibnamefont
  {Friedel}}, \bibinfo {author} {\bibfnamefont {M.~S.}\ \bibnamefont
  {Hybertsen}}, \ and\ \bibinfo {author} {\bibfnamefont {M.}~\bibnamefont
  {Schl\"uter}},\ }\href {\doibase 10.1103/PhysRevB.39.7974} {\bibfield
  {journal} {\bibinfo  {journal} {Phys. Rev. B}\ }\textbf {\bibinfo {volume}
  {39}},\ \bibinfo {pages} {7974} (\bibinfo {year} {1989})}\BibitemShut
  {NoStop}%
\bibitem [{\citenamefont {Bulutay}(2007)}]{bulutay2007s}%
  \BibitemOpen
  \bibfield  {author} {\bibinfo {author} {\bibfnamefont {C.}~\bibnamefont
  {Bulutay}},\ }\href {\doibase 10.1103/PhysRevB.76.205321} {\bibfield
  {journal} {\bibinfo  {journal} {Phys. Rev. B}\ }\textbf {\bibinfo {volume}
  {76}},\ \bibinfo {pages} {205321} (\bibinfo {year} {2007})}\BibitemShut
  {NoStop}%
\bibitem [{\citenamefont {Keister}\ \emph {et~al.}(1999)\citenamefont
  {Keister}, \citenamefont {Rowe}, \citenamefont {Kolodziej}, \citenamefont
  {Niimi}, \citenamefont {Madey},\ and\ \citenamefont
  {Lucovsky}}]{keister1999s}%
  \BibitemOpen
  \bibfield  {author} {\bibinfo {author} {\bibfnamefont {J.~W.}\ \bibnamefont
  {Keister}}, \bibinfo {author} {\bibfnamefont {J.~E.}\ \bibnamefont {Rowe}},
  \bibinfo {author} {\bibfnamefont {J.~J.}\ \bibnamefont {Kolodziej}}, \bibinfo
  {author} {\bibfnamefont {H.}~\bibnamefont {Niimi}}, \bibinfo {author}
  {\bibfnamefont {T.~E.}\ \bibnamefont {Madey}}, \ and\ \bibinfo {author}
  {\bibfnamefont {G.}~\bibnamefont {Lucovsky}},\ }\href {\doibase
  10.1116/1.590834} {\bibfield  {journal} {\bibinfo  {journal} {J. Vac. Sci.
  Technol. B}\ }\textbf {\bibinfo {volume} {17}},\ \bibinfo {pages} {1831}
  (\bibinfo {year} {1999})}\BibitemShut {NoStop}%
\bibitem [{\citenamefont {Himpsel}\ \emph {et~al.}(1988)\citenamefont
  {Himpsel}, \citenamefont {McFeely}, \citenamefont {Taleb-Ibrahimi},
  \citenamefont {Yarmoff},\ and\ \citenamefont {Hollinger}}]{himpsel1988s}%
  \BibitemOpen
  \bibfield  {author} {\bibinfo {author} {\bibfnamefont {F.~J.}\ \bibnamefont
  {Himpsel}}, \bibinfo {author} {\bibfnamefont {F.~R.}\ \bibnamefont
  {McFeely}}, \bibinfo {author} {\bibfnamefont {A.}~\bibnamefont
  {Taleb-Ibrahimi}}, \bibinfo {author} {\bibfnamefont {J.~A.}\ \bibnamefont
  {Yarmoff}}, \ and\ \bibinfo {author} {\bibfnamefont {G.}~\bibnamefont
  {Hollinger}},\ }\href {\doibase 10.1103/PhysRevB.38.6084} {\bibfield
  {journal} {\bibinfo  {journal} {Phys. Rev. B}\ }\textbf {\bibinfo {volume}
  {38}},\ \bibinfo {pages} {6084} (\bibinfo {year} {1988})}\BibitemShut
  {NoStop}%
\bibitem [{\citenamefont {Wang}, \citenamefont {Franceschetti},\ and\
  \citenamefont {Zunger}(1997)}]{lcbb1997s}%
  \BibitemOpen
  \bibfield  {author} {\bibinfo {author} {\bibfnamefont {L.-W.}\ \bibnamefont
  {Wang}}, \bibinfo {author} {\bibfnamefont {A.}~\bibnamefont {Franceschetti}},
  \ and\ \bibinfo {author} {\bibfnamefont {A.}~\bibnamefont {Zunger}},\ }\href
  {\doibase 10.1103/PhysRevLett.78.2819} {\bibfield  {journal} {\bibinfo
  {journal} {Phys. Rev. Lett.}\ }\textbf {\bibinfo {volume} {78}},\ \bibinfo
  {pages} {2819} (\bibinfo {year} {1997})}\BibitemShut {NoStop}%
\bibitem [{\citenamefont {Wang}\ and\ \citenamefont {Zunger}(1999)}]{lcbb1999s}%
  \BibitemOpen
  \bibfield  {author} {\bibinfo {author} {\bibfnamefont {L.-W.}\ \bibnamefont
  {Wang}}\ and\ \bibinfo {author} {\bibfnamefont {A.}~\bibnamefont {Zunger}},\
  }\href {\doibase 10.1103/PhysRevB.59.15806} {\bibfield  {journal} {\bibinfo
  {journal} {Phys. Rev. B}\ }\textbf {\bibinfo {volume} {59}},\ \bibinfo
  {pages} {15806} (\bibinfo {year} {1999})}\BibitemShut {NoStop}%
\bibitem [{\citenamefont {Abramowitz}\ and\ \citenamefont
  {Stegun}(1965)}]{abramowitz1965s}%
  \BibitemOpen
  \bibinfo {editor} {\bibfnamefont {M.}~\bibnamefont {Abramowitz}}\ and\
  \bibinfo {editor} {\bibfnamefont {I.~A.}\ \bibnamefont {Stegun}},\ eds.,\
  \href@noop {} {\emph {\bibinfo {title} {Handbook of Mathematical
  Functions}}}\ (\bibinfo  {publisher} {Dover Publications},\ \bibinfo
  {address} {New York},\ \bibinfo {year} {1965})\BibitemShut {NoStop}%
\end{thebibliography}
\end{document}